\documentclass[11pt]{article}

\usepackage[margin=0.75in,top=0.75in,bottom=0.75in]{geometry}
\usepackage{times}
\usepackage{microtype}
\usepackage{graphicx}
\usepackage{booktabs}
\usepackage{tabularx}
\usepackage{amsmath}
\usepackage{xcolor}
\usepackage{hyperref}
\usepackage{url}
\usepackage{multirow}
\usepackage{enumitem}
\usepackage{balance}
\usepackage{titlesec}
\usepackage{caption}
\usepackage{subcaption}
\usepackage{pifont}
\usepackage{float}

\definecolor{crit}{RGB}{185,28,28}
\definecolor{high}{RGB}{180,90,0}
\definecolor{med}{RGB}{130,100,0}
\definecolor{susc}{RGB}{202,138,4}   
\definecolor{lnkblue}{RGB}{29,78,216}

\hypersetup{colorlinks=true,linkcolor=lnkblue,citecolor=lnkblue,urlcolor=lnkblue}

\titlespacing*{\section}{0pt}{5pt}{2pt}
\titlespacing*{\subsection}{0pt}{4pt}{1pt}
\newcommand{\vc}[1]{\textcolor{crit}{\textbf{#1}}}
\newcommand{\vh}[1]{\textcolor{high}{\textbf{#1}}}
\newcommand{\vm}[1]{\textcolor{med}{\textbf{#1}}}
\newcommand{\vs}[1]{\textcolor{susc}{\textbf{#1}}}
\newcommand{\svb}{\textsc{SkillVetBench}}

\usepackage{listings}
\definecolor{promptkw}{RGB}{29,78,216}    
\definecolor{promptenum}{RGB}{22,128,90}  
\definecolor{promptcmt}{RGB}{120,113,108} 
\definecolor{promptbg}{RGB}{248,249,251}  

\lstdefinestyle{prompt}{
  basicstyle=\footnotesize\ttfamily,
  breaklines=true,
  columns=fullflexible,
  frame=single,
  rulecolor=\color{lnkblue!40},
  backgroundcolor=\color{promptbg},
  keywordstyle=\color{promptkw}\bfseries,
  commentstyle=\color{promptcmt}\itshape,
  morekeywords={ROLE,TASK,RUBRIC,SARS,CVSS,ARTIFACT,OUTPUT},
  morecomment=[s][\color{promptcmt}\itshape]{(}{)},
  emph={N,P,L,H,A,U,X,CRITICAL,HIGH,MEDIUM,LOW,NONE,INFO,true,false},
  emphstyle=\color{promptenum}\bfseries,
}

\newcommand{\sburl}{https://huggingface.co/spaces/supreme-lab/AgentSkillBench}

\begin{document}

\title{\textbf{SkillVetBench: LLM-as-Judge for Multi-Dimensional Security Risk Evaluation in Open-Source LLM Agent Skills}}

\author{%
Ismail Hossain\thanks{\texttt{ihossain@utep.edu}}, Sai Puppala, Md Jahangir Alam, Tanzim Ahad, and Sajedul Talukder\\[1pt]
\small SUPREME Lab, University of Texas at El Paso, Texas, USA.\\
\small \url{https://huggingface.co/spaces/supreme-lab/AgentSkillBench}
}
\date{}
\maketitle
\thispagestyle{empty}

\begin{abstract}
Open-source LLM agent ecosystems are growing rapidly, yet the security of
community-contributed \emph{skills} - modular tool definitions that extend agent
capabilities - remains largely unvetted.
\textbf{The gap we fill:} existing scanners operate at the code layer and are
structurally blind to \emph{instruction-layer} and \emph{multi-agent} risk -
natural-language directives that hijack an agent, exfiltrate data through encoded
side channels, or chain harm across pipelines - so what is needed is a
\emph{semantic}, multi-dimensional vetting system rather than another
signature matcher.
We present \svb{}, a live public leaderboard on Hugging Face that uses an
LLM-as-Judge to vet agent skills.
\emph{What is new:} \textbf{SARS} (Skill Agentic Risk Score), a five-dimensional
agentic-risk metric with a principled weighted formula for instruction-following
systems.
\emph{What is integrated:} full \textbf{CVSS v4.0} vector decomposition and a
\textbf{ClawHub} dual-view that places our LLM-generated review beside the
official marketplace verdict.
\emph{What is demonstrated:} drawing on our companion benchmark
paper~\cite{hossain2026svb}, the LLM-as-Judge stage achieves \textbf{zero false
negatives} across 78 confirmed-malicious skills and zero false positives across
22 benign controls, while the best static baseline (\textsc{SkillSieve}) still
misses 15\%; for instruction-layer categories such as Prompt Injection and
Memory Poisoning, conventional tools miss between 89\% and 100\% of threats
(e.g., \textsc{CodeBERT} detects none of nine memory-poisoning skills).
Detection rate varies from 35\% to 95\% across four LLM evaluators, motivating
ensemble scoring for production deployments.
All leaderboard statistics in this paper are anchored to a single snapshot
(Section~\ref{sec:stats}).
\end{abstract}

\section{Introduction}

Modern LLM agents are no longer self-contained models - they operate within
ecosystems. Frameworks such as LangChain~\cite{langchain}, Auto-GPT~\cite{autogpt},
and OpenAI's function-calling interface let agents discover, install, and invoke
community-contributed \emph{skills}: modular tool definitions that extend an
agent's executable action space at runtime. Open marketplaces such as ClawHub and
OpenClaw now host tens of thousands of these skills, illustrating both the scale
and the practical importance of the emerging agent-skill ecosystem.

The same openness that fuels this rapid capability expansion also introduces a new
supply-chain attack surface. A malicious contributor can embed harmful behavior
inside a skill that appears benign to users, agents, and marketplace scanners
alike. Unlike conventional software vulnerabilities, the danger in an agent skill
is often encoded not in inspectable executable code but in the \emph{instruction
layer} - the natural-language directives that steer the agent to execute
unauthorized shell commands, exfiltrate data through encoded side channels, poison
persistent memory, or amplify harm by chaining across multi-agent pipelines. This
risk is no longer hypothetical: recent supply-chain campaigns have introduced
large numbers of malicious skills into live marketplaces despite the presence of
official submission-vetting tools, weaponizing credential theft and data
exfiltration that surface only at execution time~\cite{hossain2026svb}.

The core difficulty is that existing defenses are structurally mismatched to this
threat. Signature-based scanners and static analyzers operate at the code layer
and are blind to instruction-layer attacks that carry no code-level signature;
ad-hoc human review by platform curators does not scale to thousands of skills;
and even where risky primitives are identified statically, no static method can
confirm whether they are actually triggered, with what arguments, and to what
effect at runtime. What has been missing is a standardized, automated, and
\emph{publicly comparable} way to evaluate agent skills across multiple security
dimensions at once.

\begin{figure}[t]
\centering
\includegraphics[width=\textwidth, height=9cm]{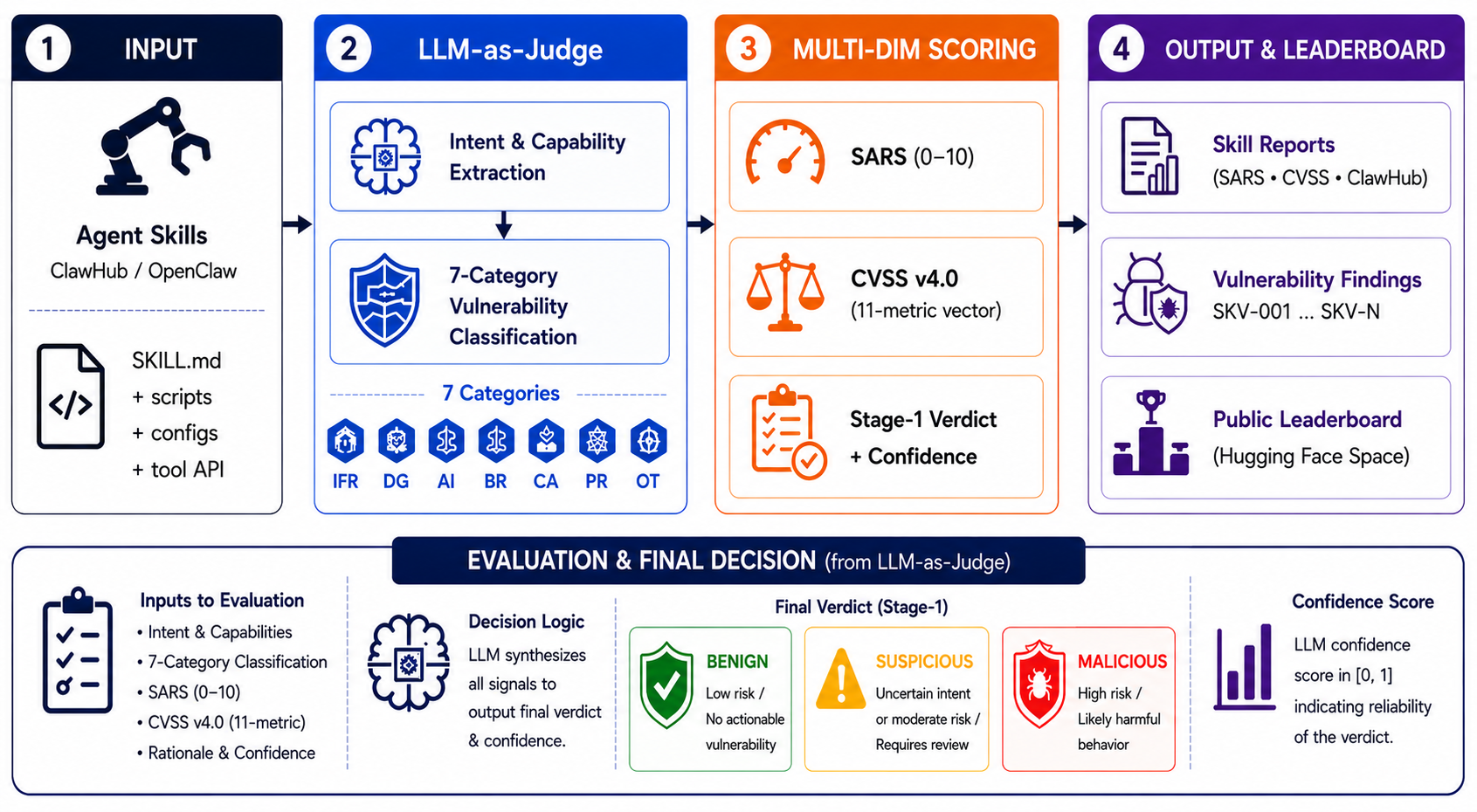}
\caption{The \svb{} system architecture. Agent skills from the
ClawHub/OpenClaw marketplace (Stage~1) are analyzed by an LLM-as-Judge that
extracts intent and capabilities and performs vulnerability classification
(Stage~2); each skill is then scored along three frameworks - SARS, CVSS v4.0,
and a Stage-1 verdict with confidence (Stage~3) - and the results are compiled
into per-skill reports, SKV-numbered findings, and the public leaderboard
(Stage~4). The bottom panel details the LLM-as-Judge decision process: it
synthesizes all extracted signals into a final three-tier verdict (Benign /
Suspicious / Malicious) accompanied by a confidence score in $[0,1]$ that
reflects the reliability of the verdict.}
\label{fig:architecture}
\end{figure}

\svb{} addresses this gap with an LLM-as-Judge that performs semantic analysis
over the full skill artifact and reports three complementary signals per skill -
the novel SARS agentic-risk score, a full CVSS~v4.0 vector, and a calibrated
three-tier verdict - consolidated into a live, public Hugging Face leaderboard. It
is the public-facing companion to the \textsc{SkillVetBench} benchmark
framework~\cite{hossain2026svb} (arXiv:2606.00925), which supplies the labeled
corpus and detection results that ground the scoring methodology. Figure~\ref{fig:architecture}
summarizes the four-stage pipeline (input $\rightarrow$ LLM-as-Judge
$\rightarrow$ multi-dimensional scoring $\rightarrow$ output/leaderboard); the
mechanics of each stage are detailed in Section~\ref{sec:framework}.

\textbf{Contributions.}
\begin{itemize}[leftmargin=1em,nosep,itemsep=0pt]
  \item \textbf{SARS} \emph{(new)}: a five-dimensional agentic-risk score with a
    principled weighted formula designed for instruction-following systems,
    capturing instruction-layer and multi-agent risk that code-level scores miss.
  \item \textbf{Multi-framework integration}: full CVSS~v4.0 vector decomposition
    per skill plus a ClawHub \emph{dual-view} that contrasts an LLM-generated
    safety review against the official marketplace verdict, surfacing detection
    gaps as a first-class signal.
  \item \textbf{A live, reproducible leaderboard} \emph{(demonstrated)}: an
    LLM-as-Judge evaluation service over a 1{,}200-skill corpus with CSV export,
    per-finding SKV records, and recorded evaluator metadata, enabling automated,
    scalable, and interpretable skill vetting.
\end{itemize}

\textbf{Roadmap.}
Section~\ref{sec:platform} describes the platform and a skill's lifecycle within
it; Section~\ref{sec:framework} details the LLM-as-Judge methodology, the SARS,
CVSS, and ClawHub components, and why they are reported jointly rather than fused;
Section~\ref{sec:taxonomy} gives the vulnerability taxonomy; Section~\ref{sec:llmjudge}
reports controlled detection results from~\cite{hossain2026svb} (a labeled
100-skill set); Section~\ref{sec:case} walks through a case study and a benign
contrast; Section~\ref{sec:stats} reports live-leaderboard statistics over 1{,}200
skills; and Section~\ref{sec:limits} states limitations. The order moves from
\emph{how the system works} to \emph{how well it works} to \emph{what it does
not yet do}.

\section{Platform Overview}
\label{sec:platform}

\begin{figure}
    \centering
    \includegraphics[width=\textwidth]{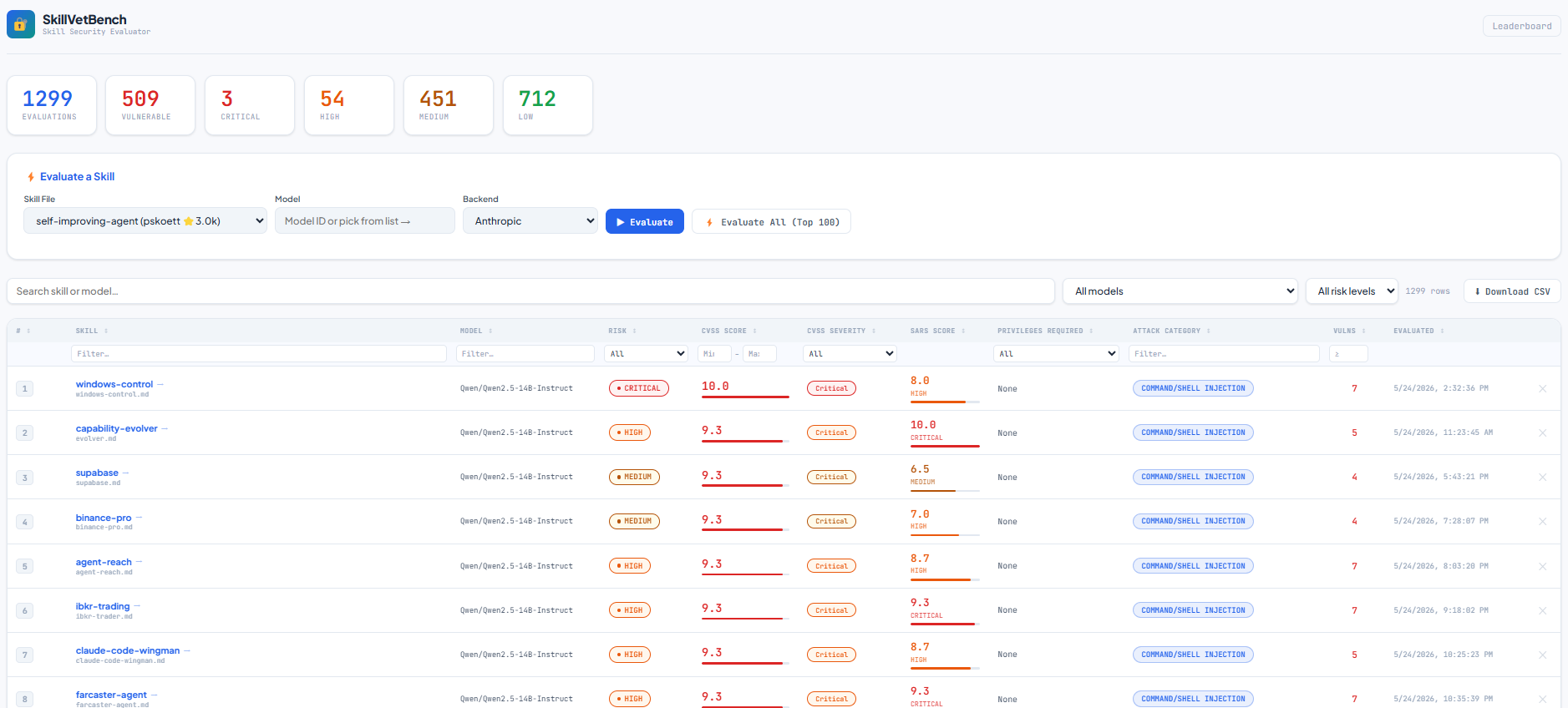}
    \caption{The leaderboard provides an overview of the evaluation results for all skills, including the total number of vulnerabilities and their distribution across Critical, High, Medium, and Low severities.}
    \label{fig:main_leaderboard}
\end{figure}

The \emph{Evaluate a Skill} panel in the Figure~\ref{fig:main_leaderboard} lets users select any skill from a dropdown
indexed by repository name and community star count (e.g.,
\texttt{self-improving-agent} with 3.0k stars), choose an evaluation model, and
select a backend. Single skills or the top-100 batch can be evaluated on demand.
The results table displays per-skill: composite \textsc{Risk} level (Critical /
High / Medium / Low), CVSS Score (0--10) with categorical severity, SARS Score
with agentic severity label, minimum Privileges Required to exploit, primary
Attack Category, count of individual vulnerability findings (\textsc{Vulns}), and
evaluation timestamp. Users can filter by any column, search by skill or model
name, and export all results as CSV for reproducible offline experiments.

\textbf{Skill lifecycle.}
A skill passes through six stages on the platform.
\emph{(1) Ingestion}: the complete artifact is fetched - its \texttt{SKILL.md}
specification together with bundled scripts, configuration files, and declared
tool interfaces.
\emph{(2) Parsing}: the artifact is decomposed into typed segments (natural-language
instructions, code blocks, config keys, tool/argument declarations) and normalized
(see Section~\ref{sec:framework} for truncation rules).
\emph{(3) Evaluation}: the LLM-as-Judge reasons over the parsed segments to extract
declared intent and capabilities and to classify findings against the taxonomy.
\emph{(4) Scoring}: the five SARS dimensions, the CVSS~v4.0 vector, and a
preliminary verdict-with-confidence are produced.
\emph{(5) Reporting}: signals are consolidated into three tabbed reports and into
per-finding SKV-numbered records.
\emph{(6) Publication}: the composite risk row is written to the public
leaderboard with the evaluating model recorded alongside the verdict.

\textbf{What the platform reads - and what it does not.}
The judge reads \emph{static} content only: the natural-language instructions, the
source of bundled scripts, configuration files, and the declared tool/argument
interfaces. It does \emph{not} execute the skill, sandbox-run its code, observe
runtime arguments, or perform dynamic taint tracking. Consequently every verdict
is an assessment of \emph{declared and inspectable} behavior; risks that
materialize only when code is run with specific inputs are inferred from static
evidence, not observed. This boundary is important for interpreting the CVSS
\emph{Exploit Maturity} field (Section~\ref{sec:cvss}) and for the evasion
limitations discussed in Section~\ref{sec:limits}.

\textbf{Deployment and refresh.}
The leaderboard is a continuously updated Hugging Face Space. New or updated
skills are scored on submission, and an evaluation is re-run when either the
underlying artifact version or the selected evaluation model changes; because the
verdict is model-dependent, each row stores the model identifier so that scores
remain comparable only within the same evaluator regime. The statistics in
Section~\ref{sec:stats} are reported against a fixed snapshot to keep the paper's
numbers reproducible despite the live system continuing to grow.

\textbf{Analyst workflow.}
A typical reviewer starts from the composite \textsc{Risk} column to triage,
sorts or filters to the High/Critical band, opens a skill's three tabbed reports
to read the SARS dimension bars, the decomposed CVSS vector, and the ClawHub
side-by-side verdicts, then drills into the individual SKV findings - each of
which carries affected content, a danger explanation, an attack scenario, and a
remediation - before exporting the filtered set to CSV for offline analysis or
for gating a publication pipeline.

\section{Evaluation Framework}
\label{sec:framework}

\noindent \svb{} is organized as the four-stage pipeline of
Figure~\ref{fig:architecture}: an \textbf{input} stage that ingests the complete
skill artifact, an \textbf{LLM-as-Judge} stage that performs semantic analysis, a
\textbf{multi-dimensional scoring} stage that emits SARS, a full CVSS~v4.0 vector,
and a preliminary verdict-with-confidence, and an \textbf{output} stage that
consolidates these into reports, SKV findings, and the leaderboard. Each skill
receives three structured reports, accessible via tabs.

\subsection{LLM-as-Judge Methodology}
\label{sec:judgemethod}

\textbf{Model, decoding, and context.}
The live leaderboard's default judge is \textsc{Qwen2.5-14B-Instruct}; the
controlled benchmark in Section~\ref{sec:llmjudge} additionally evaluates
\textsc{Qwen2.5-32B}, \textsc{Llama-3.1} (7B/70B), \textsc{Llama-3.2-3B-Instruct},
and \textsc{Mixtral-8x7B}. The same judge and the same prompt template are used
for all sub-tasks (intent extraction, taxonomy classification, SARS scoring, CVSS
vectoring, and the final verdict) within a single evaluation, so that all signals
for a skill are mutually consistent. Decoding uses low-temperature sampling
($T{=}0.2$; \texttt{top\_p}$=0.9$) to favor determinism; artifacts that exceed the
judge's context window are handled by the truncation policy below.

\textbf{Prompt structure.}
The judge prompt is a fixed template with four blocks: (i) a role/task preamble
defining the security-review objective and the three-tier verdict semantics; (ii)
the embedded SARS and taxonomy rubrics (the same ordinal definitions as
Table~\ref{tab:sars_rubric}); (iii) the parsed skill artifact, segment-tagged by
type (\texttt{<instructions>}, \texttt{<code>}, \texttt{<config>},
\texttt{<tools>}); and (iv) a structured-output schema instructing the judge to
return per-dimension SARS scores, per-finding SKV records, a CVSS vector, and a
verdict with a confidence in $[0,1]$. The condensed template is reproduced in
Appendix~\ref{app:repro}.

\textbf{Preprocessing, normalization, and truncation.}
Artifacts are parsed into typed segments, comments and whitespace are normalized,
and segments are concatenated in a fixed order (instructions, then code, then
config, then tool declarations). When an artifact exceeds the context budget,
code and config blocks are truncated last (instructions are preserved
preferentially, since instruction-layer intent is the primary threat class);
truncation is logged so that affected evaluations are auditable.

\textbf{Verdict decision rule and confidence.}
The final verdict is the judge's synthesis of all extracted signals into one of
\{Benign, Suspicious, Malicious\}, each with a confidence in $[0,1]$. A skill is
surfaced as \emph{vulnerable} on the leaderboard when its composite risk reaches
Medium or above (Section~\ref{sec:stats}).
We describe confidence as \emph{calibrated} in the sense that it tracks the
judge's stated certainty; we do \emph{not} yet report a reliability-diagram
calibration against empirical accuracy, and thresholds were not separately tuned
on a held-out validation split. Establishing post-hoc calibration (e.g.,
temperature scaling on the labeled set of~\cite{hossain2026svb}) is left to future
work and noted as a limitation in Section~\ref{sec:limits}.

\textbf{Why three frameworks instead of one fused score.}
We deliberately report SARS, CVSS, and the verdict separately rather than
collapsing them into a single number because they answer different questions and
disagree informatively. CVSS asks \emph{how exploitable and impactful is this as
code}; SARS asks \emph{how dangerous is this as an agentic actor} (instruction
hijackability, irreversibility, blast radius, chain amplification); and the verdict
is an interpretable triage label. A fused score would hide exactly the divergences
that carry the most signal - e.g., skills that are low-CVSS but high-SARS
(Section~\ref{sec:cvss}) or Malicious-by-LLM but Suspicious-by-platform
(Section~\ref{sec:clawhub}). Keeping the axes separate preserves auditability and
lets a reviewer distinguish code-dangerous from agent-dangerous skills.

\subsection{SARS: Skill Agentic Risk Score}
\label{sec:sars}

SARS is an original metric for agentic systems where conventional scores
miss instruction-layer and propagation risks.
It is computed from five 0--3 dimensions:

\vspace{-2pt}
\begin{equation}
\text{SARS} = \frac{2\,\textit{IFR} + 1.5\,\textit{DG}
  + 1.5\,\textit{AI} + 2\,\textit{BR} + 2\,\textit{CA}}{2.7}
\label{eq:sars}
\end{equation}

\noindent The denominator normalizes to a $0$--$10$ scale: the weights sum to
$9$, the maximum per-dimension score is $3$, so the un-normalized maximum is
$27$, and $27/2.7 = 10$.

\textbf{Why these five dimensions and these weights.}
The dimensions are chosen to span the orthogonal axes along which an
\emph{agentic} action becomes dangerous: how easily it can be steered off-task
(IFR), how sensitive the data it touches is (DG), whether its effects can be
undone (AI), how far harm propagates (BR), and whether it acts as a force
multiplier in a multi-skill chain (CA). IFR, BR, and CA carry the higher
weight ($2\times$) because they are the dimensions a code-layer scanner is least
able to see and that most directly govern agentic, instruction-driven harm;
DG and AI carry $1.5\times$ because they often co-vary with CVSS impact metrics
and would otherwise double-count code-level severity. The weights are a design
choice, not a fitted parameter; their robustness is examined next.

\textbf{Sensitivity to the weights.}
Because the SARS dimensions are positively correlated across malicious skills
(Table~\ref{tab:sarsdim}), the \emph{ranking} SARS induces is robust to moderate
re-weighting. As an illustration, the mean Memory-Poisoning profile
(IFR$=2.11$, DG$=1.56$, AI$=2.11$, BR$=2.00$, CA$=2.11$) scores $6.65$ under
Eq.~(\ref{eq:sars}) and $6.59$ under equal weights renormalized to $0$--$10$ -
a $<\!1\%$ shift - and the windows-control case (Section~\ref{sec:case}) scores
$8.0$ under both. We therefore treat SARS primarily as a \emph{ranking and triage}
signal whose ordering is stable under weight perturbation, with the categorical
labels (below) as a coarse severity overlay.

\textbf{Ranking vs.\ severity.}
SARS is reported as both, but the two roles are distinct: the continuous score is
a stable \emph{ranking} signal for triage, while the bands below provide a
\emph{severity} label for thresholding. Reviewers should rank with the continuous
value and gate with the band.

\begin{figure*}
    \centering
    \includegraphics[width=\textwidth]{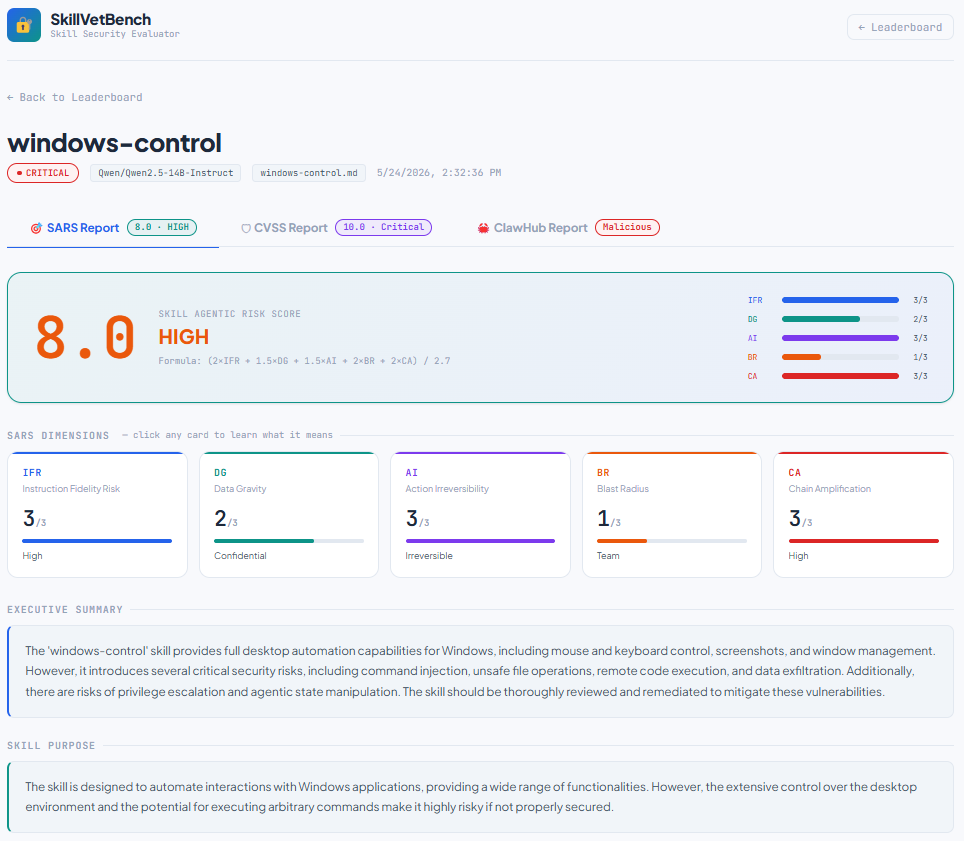}
    \caption{SARS Report for \texttt{windows-control}: composite score 8.0
    (\textsc{High}), the formula, the five dimension bars
    (IFR=3, DG=2, AI=3, BR=1, CA=3), and the executive summary. Each dimension
    bar is clickable and opens the rubric popup of Figure~\ref{fig:sars_rubric}.}
    \label{fig:sars_report}
\end{figure*}

\begin{table}[H]
\centering
\caption{
SARS dimension scoring rubrics. Each dimension is scored 0--3 by the
LLM-as-Judge over the complete skill artifact. Weights in
Eq.~(\ref{eq:sars}): IFR, BR, and CA carry $2\times$; DG and AI carry
$1.5\times$. Definitions follow Appendix~B.2 of~\cite{hossain2026svb}.}
\label{tab:sars_rubric}

\begin{tabular}{@{}p{1.8cm}c p{1.6cm} p{8.5cm}@{}}
\toprule
\textbf{Dim.} & \textbf{S} & \textbf{Level} & \textbf{Description} \\
\midrule

\multirow{4}{*}{\parbox{1.8cm}{\centering\textbf{IFR}\\Instruction\\Fidelity\\Risk}}
& 0 & Rigid
& No free-text input flows into tool behaviour; instructions are fully hardcoded. \\
& 1 & Low
& User text accepted but scoped to a fixed operation; cannot alter logic or tool selection. \\
& 2 & Medium
& User-controlled text influences API parameters or tool selection; partial hijacking surface. \\
& 3 & High
& User text incorporated directly into instructions with no sanitization; fully susceptible to prompt injection or instruction override. \\
\midrule

\multirow{4}{*}{\parbox{1.8cm}{\centering\textbf{DG}\\Data\\Gravity}}
& 0 & Public
& Only publicly available or non-sensitive data; no confidentiality risk. \\
& 1 & Internal
& Company-internal, non-sensitive data; exposure causes operational harm. \\
& 2 & Confidential
& PII, session tokens, credentials, or financial records; direct privacy or regulatory violation on exposure. \\
& 3 & Restricted
& Health records, private keys, payment instruments, or authentication secrets; severe legal or security consequences. \\
\midrule

\multirow{4}{*}{\parbox{1.8cm}{\centering\textbf{AI}\\Action\\Irreversi-\\bility}}
& 0 & Read-only
& GET-only operations; no state change; fully observable with no side effects. \\
& 1 & Reversible
& POST/PUT operations with a clear undo or rollback path. \\
& 2 & Difficult
& Shared-state modification where only partial rollback is possible. \\
& 3 & Irreversible
& Permanent actions: DELETE operations, sent messages, financial transactions, or published posts. \\
\midrule

\multirow{4}{*}{\parbox{1.8cm}{\centering\textbf{BR}\\Blast\\Radius}}
& 0 & Self
& Only the requesting user's private resources; harm fully contained to one principal. \\
& 1 & Team
& A bounded group such as a shared workspace or project unit. \\
& 2 & Platform
& All users of an integrated service or platform. \\
& 3 & Cross-sys.
& External systems, third parties, or wormable attack paths across organizational boundaries. \\
\midrule

\multirow{4}{*}{\parbox{1.8cm}{\centering\textbf{CA}\\Chain\\Amplifi-\\cation}}
& 0 & None
& Self-contained; chaining with other skills does not increase danger. \\
& 1 & Low
& Chaining adds only marginal capability; incremental risk is negligible. \\
& 2 & Medium
& Chaining creates meaningful attack paths, e.g., read-then-exfiltrate or fetch-then-execute. \\
& 3 & High
& Force multiplier: enables exfiltration, lateral movement, or persistence via execute-then-persist or spawn-then-escalate chains. \\
\bottomrule
\end{tabular}
\end{table}

\begin{figure*}
    \centering
    \includegraphics[width=\textwidth]{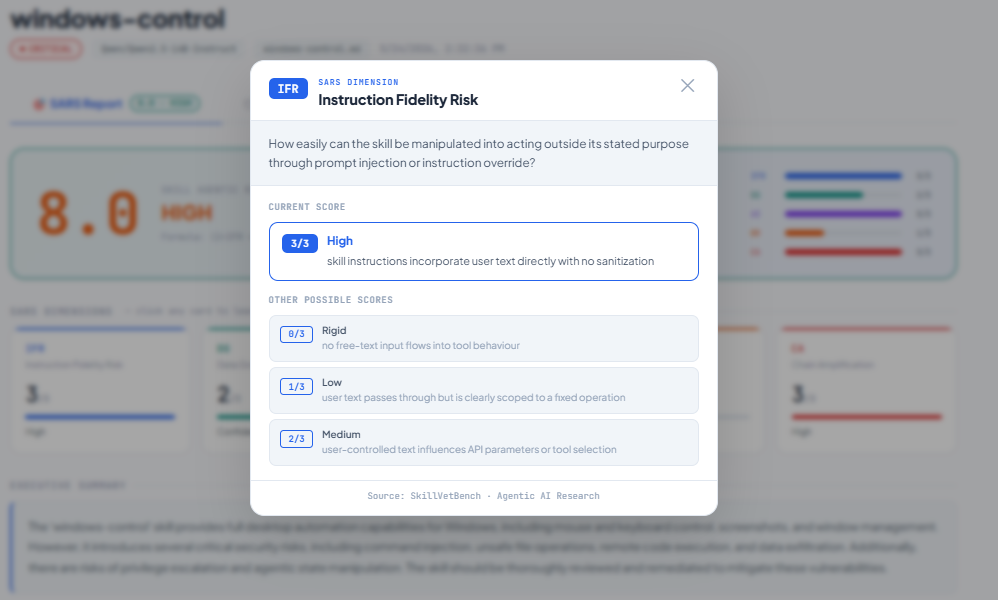}
    \caption{IFR rubric popup: the four ordinal levels (0--3) with semantic
    descriptions and the currently assigned level highlighted, making each SARS
    score assignment auditable.}
    \label{fig:sars_rubric}
\end{figure*}

Each dimension card is interactive: clicking it opens a rubric panel
(Figure~\ref{fig:sars_rubric}) showing all four ordinal levels with semantic
descriptions, enabling full auditability of score assignment.
Output labels: \vc{Critical} ($\geq$9.0), \vh{High} (7.0--8.9),
\vm{Medium} (4.0--6.9), Low ($<$4.0).

\textbf{Worked example (illustrative).}
Consider a hypothetical \texttt{inbox-cleaner} skill whose \texttt{SKILL.md}
reads: ``\emph{Given the user's request, search their mailbox and delete matching
messages.}'' Tracing the artifact to a score: IFR$=3$ because the user's free
text flows unsanitized into the search/delete instruction (instruction override
surface); DG$=2$ because mailboxes hold PII and session-bearing content; AI$=3$
because deletion is irreversible; BR$=1$ because harm is bounded to the
requesting user's mailbox; CA$=2$ because a read-then-delete pattern composes with
exfiltration skills. Eq.~(\ref{eq:sars}) gives
$(2{\cdot}3 + 1.5{\cdot}2 + 1.5{\cdot}3 + 2{\cdot}1 + 2{\cdot}2)/2.7 =
(6+3+4.5+2+4)/2.7 = 19.5/2.7 \approx 7.2$, i.e.\ \vh{High}. The example shows how
each ordinal judgment is independently auditable and how the formula composes them
into a final label.

\subsection{CVSS v4.0}
\label{sec:cvss}
\svb{} computes a full CVSS v4.0 base score~\cite{cvssv4} for each skill,
with the complete metric vector string (e.g., \texttt{CVSS:4.0/AV:N/AC:L/AT:N/PR:N/UI:N/VC:H/VI:H/VA:H/SC:H/SI:H/SA:H/E:A}).
CVSS v4.0 groups metrics into \emph{Exploitability} (Attack Requirements,
Privileges Required, User Interaction), \emph{Vulnerable System Impact}
(Confidentiality, Integrity, Availability on the directly affected system),
\emph{Subsequent System Impact} (cascading impacts beyond the exploited
system), and \emph{Threat} (Exploit Maturity).
Each metric cell is interactive, providing the rationale behind the scoring
decision.

\textbf{Why CVSS as well as SARS.}
CVSS contributes an internationally recognized, externally comparable severity
standard that security teams already operationalize in tickets, SLAs, and gating
policies; SARS contributes the agentic dimensions CVSS cannot express. Reporting
both lets \svb{} speak the language of conventional vulnerability management while
still capturing instruction-layer risk - and, crucially, the \emph{gap} between
the two is itself a signal (below).

\textbf{What the judge populates vs.\ what is derived by rule.}
The LLM-as-Judge assigns the categorical \emph{metric values} (e.g., AV:N, PR:N,
VC:H) from its semantic reading of the artifact, together with a per-metric
rationale. The numeric base score and severity band are then computed
\emph{deterministically} from that vector by the standard CVSS~v4.0 scoring
function~\cite{cvssv4}, not by the model.
This separation means readers should not assume the number is produced ``the
CVSS way'' end-to-end: the \emph{vector} is model-judged, the \emph{arithmetic}
is standard.

\textbf{Limits of CVSS on instruction-layer threats.}
CVSS was designed for code-level vulnerabilities and does not cleanly model
instruction-layer or multi-agent risk: there is no native metric for prompt
hijackability, for chain amplification across skills, or for instruction-scoped
data exfiltration. As a result, threats whose harm lives entirely in
natural-language instructions can receive a low CVSS base score even when they are
agentically dangerous - which is precisely the divergence SARS is built to expose
and which motivates reporting both.

\textbf{Low under CVSS, high under SARS.}
Table~\ref{tab:sarsdim} makes this concrete: \emph{Data Exposure} (CVSS $1.84$)
and \emph{Supply Chain} (CVSS $2.30$) both fall below the Medium CVSS band, yet
each carries elevated SARS dimensions (CA $\geq 1.80$, IFR $=2.00$) and receives a
\emph{Suspicious} verdict, because their danger is an instruction-level
exfiltration path or a runtime dependency injection that static scoring cannot
see. The same skill that looks benign as code can be dangerous as an agentic
actor.

\subsection{ClawHub Dual-View Integration}
\label{sec:clawhub}

The ClawHub report provides two sub-views.

\textbf{Why a dual view.}
Pairing our LLM-generated review with the official marketplace verdict converts a
single opaque label into a \emph{comparison}. The methodological purpose is to
make under-detection by production marketplace vetting measurable: where the two
views agree, confidence is reinforced; where they diverge, the gap localizes
exactly which security dimension the marketplace review missed. We therefore treat
disagreement not as an error to be reconciled but as a \emph{research finding and
an actionable signal} - a per-dimension map of where automated platform vetting
and deeper semantic analysis part ways.

\textbf{How the official review is obtained and aligned.}
The official sub-view surfaces the OpenClaw Safety Review published on the ClawHub
platform for the same skill. To enable a like-for-like comparison, our LLM Report
mirrors the official evaluation format and renders both reviews in the same
card-based layout over the \emph{same five dimensions} - Purpose \& Capability,
Instruction Scope, Install Mechanism, Credentials, Persistence \& Privilege - so
the comparison protocol is dimension-aligned by construction. One rubric mismatch
remains and is reported transparently: the LLM Report uses
Pass/Warning/Fail markers, whereas the official review uses an advisory scale
(e.g., \texttt{note} vs.\ \texttt{concern}); we map these onto a common
ordinal axis for side-by-side reading rather than forcing identical labels.

\begin{figure}
    \centering
    \includegraphics[width=\textwidth]{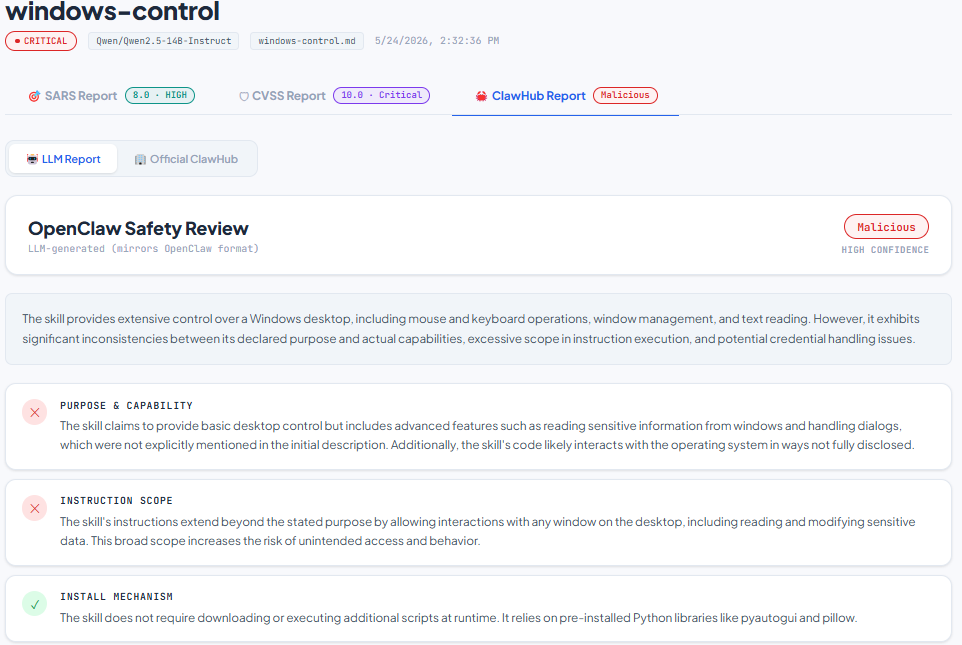}
    \caption{LLM-generated OpenClaw Safety Review for the \texttt{windows-control}
    skill, mirroring the official platform format: top-level verdict
    (\textsc{Malicious}, High Confidence), executive summary, and per-dimension
    Pass/Warning/Fail finding cards. Compare against the official review of the
    same skill (Section~\ref{sec:clawhub}).}
    \label{fig:our-llm-report}
\end{figure}

\medskip
\noindent\textbf{LLM Report.}
As illustrated in Figure~\ref{fig:our-llm-report} for the \texttt{windows-control}
skill, the report opens with a top-level verdict - here \vc{Malicious}
at \textsc{High Confidence} - followed by a concise executive summary that
synthesizes the skill's overall risk posture (flagging inconsistencies between
declared and actual capabilities, excessive instruction scope, and potential
credential-handling issues).
The body then decomposes the assessment into the five security dimensions, each
rendered as an independent finding card carrying a Pass~(\ding{51}),
Fail~(\ding{55}), or Warning~(\ding{115}) marker together with a natural-language
justification:
\begin{itemize}[leftmargin=1.2em,nosep,itemsep=1pt]
  \item \textbf{Purpose \& Capability}: whether the skill's advertised purpose
    matches its actual capabilities. For windows-control this fails
    (\ding{55}), because the skill claims to offer basic desktop control yet
    includes undisclosed advanced features such as reading sensitive on-screen
    information and handling dialogs.
  \item \textbf{Instruction Scope}: whether the skill's instructions remain
    bounded to the stated task. This also fails (\ding{55}), as the instructions
    extend to interacting with any window on the desktop - including reading and
    modifying sensitive data - broadening the attack surface well beyond the
    declared scope.
  \item \textbf{Install Mechanism}: whether the skill downloads or executes
    additional code at runtime. This passes (\ding{51}): the skill requires no
    runtime downloads and relies only on pre-installed Python libraries such as
    \texttt{pyautogui} and \texttt{pillow}.
  \item \textbf{Credentials}: whether the skill accesses or risks exposing
    sensitive credentials.
  \item \textbf{Persistence \& Privilege}: whether the skill creates persistent
    side effects or elevates privileges beyond its declared workspace.
\end{itemize}
This card-based decomposition makes the verdict auditable rather than opaque:
every contributing dimension is independently scored and justified, so a
reviewer can trace exactly which security properties drove the final
classification.

\medskip
\noindent\textbf{Official ClawHub.}
The second sub-view surfaces the \emph{official} OpenClaw Safety Review published
on ClawHub for the same skill, in the same card-based layout. 
This view first displays the skill's marketplace metadata - public popularity
(28 stars, 6.6k interactions), install count (42 installed), and version
(v1.0.0) - anchoring the assessment to a concrete, deployed artifact. The review
header then reports the platform's own verdict, \vs{Suspicious} at
\textsc{Medium Confidence}, decomposed across the same five dimensions, each with a
rationale and a per-dimension rating drawn from the official rubric (e.g.,
\texttt{note} for an advisory observation and \texttt{concern} for an elevated
risk). For instance, Instruction Scope is rated \texttt{concern} because the
skill's defaults permit actions across any or all visible windows and dialogs
without documented approval, app allowlisting, or destructive-action safeguards,
while Persistence \& Privilege is rated \texttt{note} since the skill inherits
only the access the logged-in user already holds and adds no background
persistence, network exfiltration, or credential storage.

The central analytical value of presenting both sub-views lies in the
\emph{disagreement} they expose. For windows-control, the LLM Report
returns \vc{Malicious} at High Confidence, whereas the official ClawHub review
returns only \vs{Suspicious} at Medium Confidence - despite the skill being
actively deployed across 42 installations. By rendering both verdicts in a unified,
dimension-aligned interface, \svb{} turns marketplace under-detection into a
first-class, auditable signal.

\section{Vulnerability Taxonomy}
\label{sec:taxonomy}

\svb{} identifies vulnerabilities under \textbf{seven} categories
(Table~\ref{tab:taxonomy}). Each instance receives a unique \textbf{SKV-XXX}
identifier and is documented with: the \emph{affected content} (code or
instruction snippet), a \emph{danger explanation}, a \emph{concrete attack
scenario}, and a \emph{remediation recommendation}.

\textbf{Multiple categories and the dominant label.}
A single skill frequently exhibits findings in several categories (the
windows-control case in Section~\ref{sec:case} carries seven SKV findings across
six categories). Every finding is recorded individually; the leaderboard's
\textsc{Attack Category} column reports the \emph{dominant} category, defined as
the one containing the highest-severity finding (ties broken by finding count,
then by category order). The fine-grained leaderboard breakdown in
Table~\ref{tab:attack_cat} therefore reflects dominant categories, and groups
closely related types (e.g., Command/Shell Injection with RCE) for reporting.

\begin{table}[H]
\centering
\caption{Seven-category vulnerability taxonomy in \svb{}, with formal
definitions and distinguishing boundary notes.}
\label{tab:taxonomy}
\vspace{-2pt}
\small
\setlength{\tabcolsep}{3pt}
\begin{tabularx}{\columnwidth}{lX}
\toprule
\textbf{Category} & \textbf{Definition \& boundary} \\
\midrule
Command/Shell Injection & Shell commands whose structure or arguments are
influenced by user-controlled input, enabling arbitrary OS command execution.
\emph{Boundary:} distinguished from RCE by acting through a shell interpreter
rather than a language \texttt{eval}. \\
Remote Code Execution & Dynamic evaluation of attacker-influenced strings (e.g.,
\texttt{eval()}, dynamic import) enabling arbitrary in-process execution.
\emph{Boundary:} no shell needed; the host language runs the payload. \\
Unsafe File Operations & Attacker-influenced paths used to read, overwrite, or
delete files outside the declared workspace. \emph{Boundary:} concerns the
\emph{filesystem target}, even if no shell or eval is involved. \\
Prompt Injection & Natural-language input that overrides or rewrites the skill's
instructions to make the agent act outside its stated purpose.
\emph{Boundary:} the malicious effect is \emph{transient} (this turn) and lives
in instructions, not persisted state - this is what separates it from Memory
Poisoning. \\
Memory Poisoning & Persistent writes to long-term agent memory/state that
redirect \emph{future} behavior. \emph{Boundary:} unlike Prompt Injection, the
effect \emph{outlives} the turn; unlike Agentic State Manipulation, it targets
durable memory rather than in-flight goals. \\
Data Exfiltration & Sensitive data encoded (e.g., Base64) and transmitted through
a covert or undeclared side channel. \emph{Boundary:} requires an \emph{egress}
path; mere access to sensitive data without exfiltration is scored via DG, not
here. \\
Privilege Escalation & \texttt{sudo}/admin or capability-elevating actions beyond
the skill's declared scope. \emph{Boundary:} concerns \emph{authority}, not data
or code-injection per se. \\
\bottomrule
\end{tabularx}
\end{table}


\textbf{Taxonomy validation.}
The seven categories were derived from, and validated against, the labeled
malicious corpus of the companion benchmark~\cite{hossain2026svb}: every
confirmed-malicious skill maps to at least one category, and the category
assignments underlie the per-category detection results reported in
Section~\ref{sec:llmjudge}.

Reports also contrast \emph{dangerous patterns}
(\texttt{os.system()}, \texttt{subprocess.call()}, \texttt{eval()},
\texttt{sudo}, base64 encoding, admin invocations) against observed
\emph{safe practices} (e.g., FAILSAFE flags, input validation).
A \textsc{Remediation Priority} statement directs developers to address the
highest-severity finding class first.

\section{LLM-as-Judge: Semantic Analysis Results}
\label{sec:llmjudge}

The results in this section come from the \emph{controlled, labeled} evaluation of
the companion benchmark~\cite{hossain2026svb}: \textbf{78 confirmed-malicious
skills and 22 benign controls} (100 total). This is a different population from
the live-leaderboard corpus analyzed in Section~\ref{sec:stats} (1{,}299 skills),
and the two should not be conflated; detection metrics (Tables~\ref{tab:detection}--\ref{tab:sarsdim})
are reported only on the labeled set, where ground truth exists.

\subsection{Detection Performance vs.\ Baselines}

Table~\ref{tab:detection} compares \svb{}'s LLM-as-Judge against the
strongest baselines on overall catch rate (recall) and miss rate (FNR).
\svb{} is the only system to achieve zero false negatives across all 78
skills and zero false positives across all 22 benign controls.

\begin{table}[H]
\centering
\caption{Detection performance on 78 malicious + 22 benign skills.
Miss Rate (lower is better) is the primary safety metric.
Bold = best per column. Results from~\cite{hossain2026svb}.}
\label{tab:detection}
\vspace{-2pt}
\small
\setlength{\tabcolsep}{3.5pt}
\begin{tabular}{lcccc}
\toprule
\textbf{System} & \textbf{Balance} & \textbf{Catch} & \textbf{Precision} & \textbf{Miss\,\%} \\
\midrule
VirusTotal       & 0.46 & 0.33 & 1.00 & 67\% \\
ClawScan         & 0.56 & 0.48 & 1.00 & 52\% \\
ClawVet          & 0.53 & 0.41 & 1.00 & 59\% \\
LLM (0-shot)     & 0.76 & 0.74 & 0.87 & 26\% \\
LLM (few-shot)   & 0.80 & 0.78 & 0.88 & 22\% \\
CodeBERT         & 0.68 & 0.70 & 0.95 & 30\% \\
SkillProbe       & 0.82 & 0.81 & 0.88 & 19\% \\
SkillSieve       & 0.84 & 0.85 & 0.90 & 15\% \\
\midrule
\textbf{\svb{} (Ours)} & \textbf{0.95} & \textbf{1.00} & \textbf{0.96} & \textbf{0\%} \\
\bottomrule
\end{tabular}
\end{table}


\textbf{Confusion matrix and statistical reliability.}
On the labeled set, \svb{}'s stated outcome is fully determined: TP$=78$, FN$=0$,
FP$=0$, TN$=22$ (Table~\ref{tab:confusion}). Because the sets are small, we report
distribution-free Wilson 95\% intervals so the ``zero-miss'' claim is not
over-read: with $78/78$ recall the lower bound is $\approx 0.95$ (true miss rate
could be up to $\sim$5\%), and with $22/22$ specificity the lower bound is
$\approx 0.85$. The point estimates are perfect; the intervals reflect that 100
labeled skills cannot certify perfection, and they directly motivate the larger,
multi-judge evaluation that follows.

\begin{table}[H]
\centering
\caption{Confusion matrix for \svb{} on the labeled 100-skill set, with
Wilson 95\% lower bounds for the small-sample point estimates.}
\label{tab:confusion}
\small
\setlength{\tabcolsep}{6pt}
\begin{tabular}{lcc}
\toprule
 & \textbf{Pred.\ Malicious} & \textbf{Pred.\ Benign} \\
\midrule
\textbf{Actual Malicious} (78) & 78 (TP) & 0 (FN) \\
\textbf{Actual Benign} (22)    & 0 (FP)  & 22 (TN) \\
\midrule
\multicolumn{3}{l}{\footnotesize Recall $=1.00$ (95\% CI lower $\approx 0.95$);
Specificity $=1.00$ (lower $\approx 0.85$).} \\
\bottomrule
\end{tabular}
\end{table}

\textbf{Instruction-layer vs.\ code-layer threats.}
The detection gap is sharpest on \emph{instruction-layer} categories, whose
malicious intent has no reliable code-level signature. For \textbf{Prompt
Injection} (19 skills), \textsc{VirusTotal} flags 0 and \textsc{ClawScan} flags
only 3, while the LLM-as-Judge flags all 19. For \textbf{Memory Poisoning} (9
skills), \textsc{ClawScan} detects 1, \textsc{VirusTotal} detects 3, and
\textsc{CodeBERT} detects 0; the LLM-as-Judge classifies all 9 as Suspicious or
Malicious. By contrast, \emph{code-layer} categories (e.g., Command/Shell
Injection, RCE) carry detectable primitives that static scanners partially catch,
which is why the baselines' aggregate miss rate (15--67\%) is far better than
their near-total miss on the instruction layer.

\subsection{LLM Evaluator Sensitivity}

Detection quality depends materially on the choice of LLM judge.
Table~\ref{tab:llmsens} reports vulnerable-skill detection rates and mean
SARS scores across four evaluators tested in~\cite{hossain2026svb} on the
labeled set.

\begin{table}[H]
\centering
\caption{Sensitivity to LLM evaluator choice on the labeled
set~\cite{hossain2026svb}. Detection rates range from 35\% to 95\%.
(Distinct from the live-leaderboard evaluator breakdown in
Table~\ref{tab:model_sens}, which uses a different model roster and population.)}
\label{tab:llmsens}
\vspace{-2pt}
\small
\setlength{\tabcolsep}{3pt}
\begin{tabular}{lccc}
\toprule
\textbf{LLM Judge} & \textbf{Vuln.\,\%} & \textbf{SARS (mean)} & \textbf{Profile} \\
\midrule
Qwen2.5-32B        & 95\% & 5.06\,$\pm$\,2.02 & High recall, calibrated \\
Llama-3.1-7B       & 78\% & 4.99\,$\pm$\,2.48 & Broad category coverage \\
Llama-3.2-3B-Ins   & 43\% & 5.57\,$\pm$\,2.66 & Sparse but over-sensitive \\
Mixtral-8x7B       & 35\% & 1.74\,$\pm$\,2.20 & Systematic FN bias \\
\bottomrule
\end{tabular}
\end{table}

\textsc{Mixtral-8x7B} shows near-uniform abstention (median SARS = 0.00),
while \textsc{Llama-3.2-3B-Ins} fires sparsely but with high variance,
producing a mean of 6.50\,$\pm$\,7.80 vulnerabilities per skill  -  indicating
poorly calibrated over-detection.
These results support the recommendation in~\cite{hossain2026svb} to
use evaluators of at least 7B-parameter scale with strong instruction-following
alignment, and to apply ensemble scoring across multiple judges to reduce
single-model bias.
The live leaderboard currently uses \textsc{Qwen2.5-14B-Instruct} as the
default evaluation model.

\subsection{SARS Dimensions vs.\ CVSS: Where They Diverge}

Table~\ref{tab:sarsdim} shows mean SARS dimension scores per vulnerability
category alongside the mean CVSS v4.0 base score~\cite{hossain2026svb}.

\begin{table}[H]
\centering
\caption{Mean SARS dimension scores (0--3) and mean CVSS v4.0 base score per
vulnerability category~\cite{hossain2026svb}. Bold = highest per column.
CA stays at or above 1.80 across \emph{all} malicious categories.}
\label{tab:sarsdim}
\vspace{-2pt}
\small
\setlength{\tabcolsep}{2.8pt}
\begin{tabular}{lcccccc}
\toprule
\textbf{Category} & \textbf{IFR} & \textbf{DG} & \textbf{AI} & \textbf{BR} & \textbf{CA} & \textbf{CVSS} \\
\midrule
Cmd.\ Injection   & \textbf{2.19} & 1.70 & 2.00 & 1.41 & \textbf{2.19} & 4.16 \\
Prompt Injection  & 2.00 & 1.32 & 1.79 & 1.32 & 2.00 & 3.57 \\
Unsafe File Ops   & 2.00 & 1.20 & 1.60 & 1.00 & 1.90 & 2.62 \\
Memory Poisoning  & 2.11 & 1.56 & \textbf{2.11} & \textbf{2.00} & 2.11 & \textbf{4.54} \\
Data Exposure     & 2.00 & 1.40 & 1.40 & 1.00 & 1.80 & 1.84 \\
Supply Chain      & 2.00 & 1.00 & 1.50 & 1.00 & 2.00 & 2.30 \\
Privilege Abuse   & 2.00 & 1.50 & 2.00 & 1.75 & 2.00 & 4.08 \\
No Issue          & 0.86 & 0.32 & 0.14 & 0.05 & 1.00 & 0.00 \\
\bottomrule
\end{tabular}
\end{table}

Two divergence patterns are critical.
First, \textbf{CA (Chain Amplification) stays at or above 1.80 across every
malicious category} - including benign-appearing skills - confirming that all
confirmed malicious skills contribute to compositional attack chains regardless of
their primary vector.
Second, \textbf{Data Exposure and Supply Chain score low on CVSS} (1.84 and
2.30, below Medium) yet receive \emph{Suspicious} verdicts from the
LLM-as-Judge, because their threat arises from instruction-level exfiltration
paths and runtime dependency injection that static scoring cannot model.
This divergence is exactly where the LLM-as-Judge adds irreplaceable value
over conventional vulnerability scoring.

\section{Case Study: \texttt{windows-control}}
\label{sec:case}

We walk through \texttt{windows-control} (file: \texttt{windows-control.md},
evaluated with Qwen2.5-14B-Instruct via the platform's default inference
backend, 24 May 2026) to illustrate how the three reports interact.
We chose this skill because it is \emph{representative of the high-severity band
while also exhibiting the cross-system disagreement} the platform is designed to
surface: it is an actively deployed marketplace skill (42 installs, 6.6k
interactions) whose declared purpose (basic desktop control) understates its
actual capability, and on which our LLM review and the official platform review
disagree. It is presented as an \emph{illustrative} example of the reporting
pipeline, not as a statistical sample; corpus-level distributions are in
Section~\ref{sec:stats}. The skill provides full Windows desktop automation:
mouse/keyboard control, screenshots, and window management.

\textbf{SARS: 8.0 HIGH.}
With IFR\,=\,3, DG\,=\,2, AI\,=\,3, BR\,=\,1, CA\,=\,3, Eq.~(\ref{eq:sars})
gives $(6+3+4.5+2+6)/2.7 = 21.5/2.7 \approx 8.0$ (HIGH).
IFR\,=\,3 because skill instructions incorporate user text with no
sanitization; AI\,=\,3 because desktop actions (file writes, keystrokes,
screen captures) are irreversible; CA\,=\,3 due to high multi-agent
amplification potential.
Figures~\ref{fig:sars_report} and~\ref{fig:sars_rubric} show the full SARS
dashboard and the IFR rubric.

\textbf{CVSS v4.0: 10.0 Critical.}
Vector:\\
{\small\texttt{CVSS:4.0/AV:N/AC:L/AT:N/PR:N/UI:N/VC:H/VI:H/VA:H/SC:H/SI:H/SA:H/E:A}}.
Network-accessible, no privileges or user interaction required, High C/I/A
impact on both the vulnerable and subsequent systems, Exploit Maturity =
\emph{Attacked} (active exploitation confirmed).

\begin{figure*}[ht!]
\centering
\includegraphics[width=\textwidth]{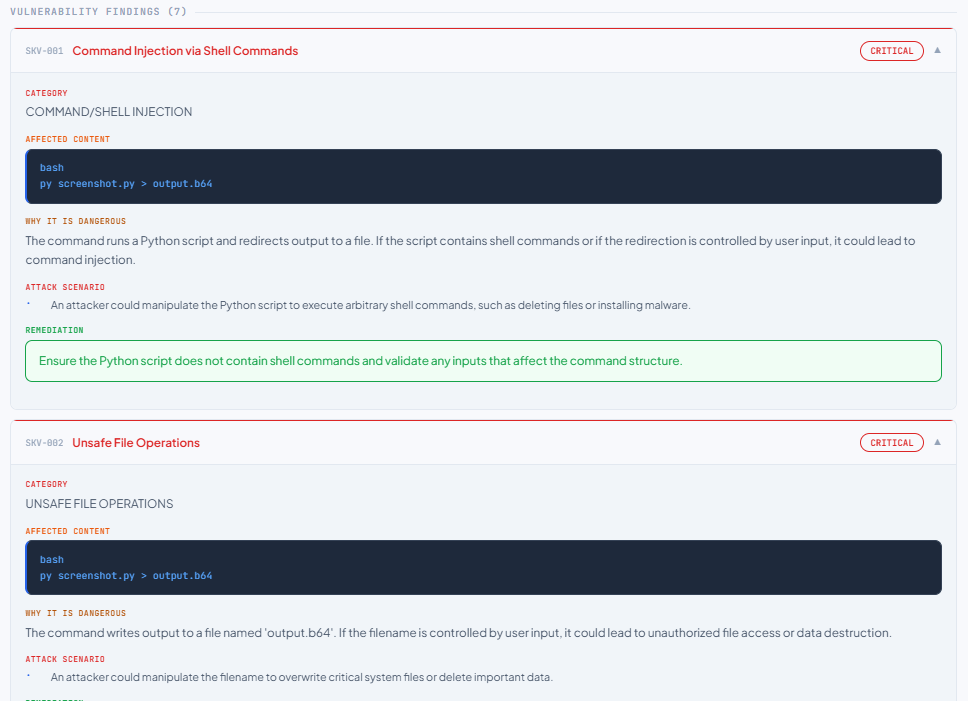}
\caption{First two structured findings (SKV-001, SKV-002) for
  \texttt{windows-control}. Each finding card reports the affected content
  (here a shell redirection of script output), why it is dangerous, a concrete
  attack scenario, and a remediation, so a developer can act on each item
  independently.}
\label{fig:vuln}
\end{figure*}

\textbf{Vulnerability Findings (7 SKV findings).}
Figure~\ref{fig:vuln} shows the first two structured findings;
all seven are:
\vc{[CRITICAL] SKV-001}  -  Command Injection via
\texttt{py screenshot.py > output.b64} (user-controlled shell redirection);
\vc{[CRITICAL] SKV-002}  -  Unsafe File Operations on the same output path;
\vc{[CRITICAL] SKV-003}  -  Remote Code Execution via \texttt{eval()};
\vh{[HIGH] SKV-004}  -  Data Exfiltration via Base64-encoded screen content;
\vh{[HIGH] SKV-005}  -  Privilege Escalation via \texttt{sudo}/admin actions;
\vh{[HIGH] SKV-006}  -  Memory Poisoning via persistent memory writes;
\vm{[MEDIUM] SKV-007}  -  Agentic State Manipulation via intermediate goal
modification.
Dangerous patterns detected: \texttt{os.system()},
\texttt{subprocess.call()}, \texttt{eval()}, base64 encoding, \texttt{sudo},
admin actions.
Safe practices also observed: \texttt{pyautogui.FAILSAFE = True}, action
delays, smooth mouse movements.

\textbf{ClawHub: Malicious vs.\ Suspicious.}
The LLM-generated OpenClaw review rates the skill \vc{Malicious}
(High Confidence), flagging: (i) inconsistency between declared purpose
(``basic desktop control'') and actual capabilities (reading credential-bearing
window content); (ii) instruction scope extending to any visible window without
approval or allowlisting.
The \emph{Official ClawHub} platform review rates the same skill
\vs{Suspicious} (Medium Confidence) - despite 6.6k interactions and 42 active
installs at v1.0.0.
This Malicious/Suspicious discrepancy (High vs.\ Medium confidence) directly
illustrates the detection gap that \svb{} is designed to surface.

\textbf{Contrast: a benign skill is not over-flagged.}
To show the system does not flag everything, consider the aggregate
\emph{No Issue} profile from Table~\ref{tab:sarsdim}
(IFR$=0.86$, DG$=0.32$, AI$=0.14$, BR$=0.05$, CA$=1.00$). Eq.~(\ref{eq:sars})
yields $(1.72+0.48+0.21+0.10+2.00)/2.7 \approx 1.67$, i.e.\ \textbf{Low}, with a
mean CVSS of $0.00$ and a \emph{Benign} verdict - read-only, self-contained skills
score near the floor on every dimension. This confirms that the high scores above
are driven by the artifact's actual risk surface rather than by a systematically
alarmist judge.

\section{Leaderboard Statistics}
\label{sec:stats}

We summarize the state of the \svb{} leaderboard across its 1{,}299 evaluations,
as of the snapshot date \textbf{24 May 2026}.
The statistics characterize the leaderboard along four axes: composite risk, the
two scoring frameworks (CVSS v4.0 and SARS), the distribution of attack
categories, and the behavior of the LLM-as-Judge evaluators that produce these
verdicts. This is the live corpus, distinct from the labeled 100-skill detection
set of Section~\ref{sec:llmjudge}.

\begin{table*}[t]
\centering
\begin{minipage}[t]{0.46\textwidth}
\centering
\caption{Composite risk-level distribution across all 1{,}299 evaluations.
Skills at Medium or above are flagged \emph{vulnerable}; 79 evaluations (6.1\%)
failed and are reported as ERROR.}
\label{tab:overall_risk}
\small
\begin{tabular}{lrr}
\toprule
\textbf{Risk Level} & \textbf{Count} & \textbf{\% of Total} \\
\midrule
\vc{Critical} &    3 & 0.2\%  \\
\vh{High}     &   54 & 4.2\%  \\
\vm{Medium}   &  451 & 34.7\% \\
Low           &  712 & 54.8\% \\
\midrule
\textbf{Vulnerable} & \textbf{509}   & \textbf{39.2\%} \\
\textbf{Total}      & \textbf{1{,}299} & \textbf{100\%} \\
\bottomrule
\end{tabular}
\end{minipage}
\hfill
\begin{minipage}[t]{0.50\textwidth}
\centering
\caption{Severity under the two scoring frameworks. CVSS v4.0 assigns a
categorical severity to the 536 scored skills; SARS assigns an agentic-risk
severity to all 1{,}299 evaluations (None = benign or unscored).}
\label{tab:severity}
\small
\setlength{\tabcolsep}{4pt}
\begin{tabular}{lrr@{\hspace{1em}}lrr}
\toprule
\multicolumn{3}{c}{\textbf{CVSS v4.0}} & \multicolumn{3}{c}{\textbf{SARS}} \\
\cmidrule(r){1-3}\cmidrule(l){4-6}
\textbf{Sev.} & \textbf{N} & \textbf{\%} & \textbf{Sev.} & \textbf{N} & \textbf{\%} \\
\midrule
\vc{Crit.} &  24 &  4.5 & \vc{Crit.} &   8 &  0.6 \\
\vh{High}  & 140 & 26.1 & \vh{High}  &  40 &  3.1 \\
\vm{Med.}  & 130 & 24.3 & \vm{Med.}  & 418 & 32.2 \\
Low        & 242 & 45.1 & Low        & 326 & 25.1 \\
\bottomrule
\end{tabular}
\end{minipage}
\end{table*}


\textbf{Composite risk (Table~\ref{tab:overall_risk}).}
Of the 1{,}299 evaluations, 509 (39.2\%) are flagged as vulnerable. The
distribution is strongly bottom-heavy: most skills fall into Low (712, 54.8\%)
and Medium (451, 34.7\%), while only 54 (4.2\%) reach High and just 3 (0.2\%)
reach Critical. This long tail is expected for a real-world marketplace - most
published skills are benign or low-risk - and the leaderboard's value lies in
reliably isolating the small but consequential high-severity minority. A further
79 evaluations (6.1\%) terminated in an ERROR state (e.g., malformed artifacts or
evaluator timeouts) and are surfaced transparently rather than silently dropped,
so the pipeline's own success rate remains auditable.

\textbf{CVSS v4.0 vs.\ SARS (Table~\ref{tab:severity}).}
The two frameworks produce markedly different severity profiles by design.
CVSS v4.0 scores only the 536 skills with a non-zero base score and concentrates
mass at the lower end (Low 45.1\%, Medium 24.3\%) while still escalating 140
(26.1\%) to High and 24 (4.5\%) to Critical, reflecting sensitivity to
network-accessible, code-level exploitability. SARS, computed over all 1{,}299
evaluations, places the bulk at Medium (418, 32.2\%) and None (507, 39.0\%), with
a deliberately narrow High/Critical band (48 skills, 3.7\% combined): it requires
multiple agentic dimensions to align before escalating, whereas CVSS can rate a
skill High on code-level impact alone. Reporting both lets a reviewer distinguish
skills dangerous \emph{as code} from skills dangerous \emph{as agentic actors}.

\begin{table*}[t]
\centering
\begin{minipage}[t]{0.48\textwidth}
\centering
\caption{The 509 vulnerable skills by top-ranked (dominant) attack category.
Command/Shell Injection and RCE dominate, accounting for nearly half of all
flagged skills. (Percentages sum to 100\% up to rounding.)}
\label{tab:attack_cat}
\small
\setlength{\tabcolsep}{4pt}
\begin{tabular}{lrr}
\toprule
\textbf{Top Attack Category} & \textbf{Count} & \textbf{\% Vuln.} \\
\midrule
Command/Shell Inj.\ \& RCE      & 236 & 46.4\% \\
Prompt Injection                &  58 & 11.4\% \\
Credential / Secret Exposure    &  51 & 10.0\% \\
Memory Poisoning \& Persist.    &  44 &  8.6\% \\
Unsafe File Operations          &  42 &  8.3\% \\
Agentic State Manipulation      &  35 &  6.9\% \\
Data Exfiltration               &  26 &  5.1\% \\
Privilege Escalation            &   9 &  1.8\% \\
Dependency / Supply Chain       &   7 &  1.4\% \\
Insecure Deserialization        &   1 &  0.2\% \\
\midrule
\textbf{Total} & \textbf{509} & \textbf{100\%} \\
\bottomrule
\end{tabular}
\end{minipage}
\hfill
\begin{minipage}[t]{0.48\textwidth}
\centering
\caption{Outcomes per LLM-as-Judge evaluator on the live leaderboard. Detection
rate varies sharply with model scale and alignment, motivating ensemble scoring.
CVSS and SARS columns report mean base scores; V/sk is the mean finding count.}
\label{tab:model_sens}
\small
\setlength{\tabcolsep}{3.5pt}
\begin{tabular}{lrrrrr}
\toprule
\textbf{LLM-as-Judge} & \textbf{N} & \textbf{Vuln\%} & \textbf{CVSS} & \textbf{SARS} & \textbf{V/sk} \\
\midrule
Qwen2.5-32B    & 120 & 79.2 & 2.97 & 5.06 & 2.48 \\
Llama-3.1-70B  & 101 & 79.2 & 3.35 & 4.90 & 4.09 \\
Mixtral-8x7B   &  99 & 35.4 & 0.52 & 1.53 & 0.94 \\
Qwen2.5-14B    & 979 & 30.5 & 1.63 & 2.12 & 1.08 \\
\bottomrule
\end{tabular}
\end{minipage}
\end{table*}


\textbf{Attack categories (Table~\ref{tab:attack_cat}).}
Among the 509 vulnerable skills, the landscape is heavily skewed toward
execution-layer risk. Command/Shell Injection and RCE together account for 236
skills (46.4\%) - nearly half - confirming that unsafe invocation of system
primitives is the single most prevalent danger in the open-skill ecosystem. The
next tier is dominated by instruction- and data-layer threats: Prompt Injection
(58, 11.4\%), Credential/Secret Exposure (51, 10.0\%), Memory Poisoning and
Persistence (44, 8.6\%), and Unsafe File Operations (42, 8.3\%). Agentic-specific
categories - Agentic State Manipulation (35, 6.9\%) and Data Exfiltration
(26, 5.1\%) - form a meaningful middle band, while Privilege Escalation (9),
Supply Chain (7), and Insecure Deserialization (1) constitute a sparse tail. An
effective vetting tool must cover both ends: the high-volume execution threats
static scanners partially catch, and the instruction-layer threats only semantic
LLM-as-Judge analysis can reliably surface.

\textbf{Evaluator sensitivity (Table~\ref{tab:model_sens}).}
The four leaderboard models split into two regimes. The larger, well-aligned
models - Qwen2.5-32B and Llama-3.1-70B - each flag $\sim$79\% of the skills they
evaluate, with the highest mean CVSS (2.97 and 3.35) and SARS (5.06 and 4.90);
Llama-3.1-70B surfaces the most findings per skill (4.09), a high-recall profile.
In contrast, Mixtral-8x7B flags only 35.4\% with a near-floor mean SARS of 1.53
(systematic false-negative bias), and the default Qwen2.5-14B - which carries the
bulk of the leaderboard at 979 evaluations - flags 30.5\%, consistent with a
smaller, more conservative judge. This nearly $2.5\times$ spread in detection rate
(35\%--79\%) is the clearest evidence that single-model verdicts are unreliable in
isolation. It motivates ensemble scoring across judges of at least moderate scale,
and explains why the leaderboard records the evaluating model alongside every
verdict so that scores remain comparable only within the same evaluator regime.

\section{Community Use Cases}

\noindent\textbf{Security researchers} can benchmark detection systems using
the seven-category taxonomy and download results via CSV for reproducible
analysis.
\textbf{Skill developers} can evaluate skills before publishing and follow
the per-SKV remediation guidance.
\textbf{AI safety researchers} can use the 1{,}299-skill corpus to study which
SARS dimensions (IFR, DG, AI, BR, CA) correlate most strongly with observed
harm.
\textbf{Platform operators} can integrate the evaluation pipeline as an
automated gate, enforcing minimum CVSS and SARS thresholds before skill
publication. Public, leaderboard-style benchmarking has driven progress in other
areas of language technology~\cite{superglue}; \svb{} aims to play the same role
for agent-skill security.

\section{Limitations}
\label{sec:limits}

\textbf{Judge dependence, drift, and nondeterminism.}
Verdicts are produced by an LLM-as-Judge, so they inherit the model's biases:
detection rate varies $35\%$--$95\%$ across judges (Tables~\ref{tab:llmsens},
\ref{tab:model_sens}), low-temperature sampling reduces but does not eliminate
run-to-run nondeterminism, and a model upgrade can shift the entire leaderboard
(model drift). We mitigate by recording the evaluator per row and recommending
ensemble scoring, but single-model rows should be read with this caveat.

\textbf{Prompt sensitivity and calibration.}
Results depend on the fixed prompt template; rubric wording and segment ordering
can affect scoring. The reported confidence reflects the judge's stated certainty
and is not yet validated against empirical accuracy via a reliability diagram;
post-hoc calibration is future work.

\textbf{Static-only analysis and adversarial evasion.}
Because \svb{} reads but never executes skills (Section~\ref{sec:platform}), it is
vulnerable to evasion strategies that a static reader can miss: instruction or
string \emph{obfuscation} (encoding, splitting, homoglyphs); \emph{multilingual}
or low-resource-language instructions that weaken the judge's detection;
\emph{indirect prompting} where malicious directives are fetched at runtime rather
than written in \texttt{SKILL.md}; \emph{wrapper code} that dynamically assembles
payloads; and benign-looking shells that activate harmful behavior only under
specific runtime inputs. A further, agent-specific risk is \emph{prompt injection
against the judge itself}: an adversarial author can embed text aimed at the
reviewing model. Hardening the judge against these attacks - and complementing
static review with sandboxed dynamic analysis - is important future work.

\textbf{Sensitivity vs.\ false alarms.}
There is an inherent tradeoff between catching subtle instruction-layer threats
and over-flagging legitimate, capability-rich automation skills (e.g., a genuine
desktop-automation tool shares primitives with a malicious one). The benign
contrast in Section~\ref{sec:case} and the bottom-heavy distribution in
Table~\ref{tab:overall_risk} suggest the default judge is conservative rather than
alarmist, but operators tightening thresholds for higher recall should expect more
false positives, and the small (22-skill) benign control set limits how tightly we
can currently bound the false-positive rate.

\textbf{Marketplace coverage and labels.}
ClawHub/OpenClaw metadata and official reviews are external inputs that may change
or be unavailable, and ``confirmed-malicious'' labels come from the companion
benchmark's labeling process~\cite{hossain2026svb}; both bound the generality of
the conclusions.

\section{Conclusion}

\svb{} is a live, public leaderboard providing multi-dimensional security
assessment of open-source LLM agent skills. Instruction-layer threats constitute a
growing attack surface that traditional scanners cannot adequately evaluate; \svb{}
shows that LLM-as-Judge semantic analysis provides a scalable mechanism for
exposing such risks and for establishing transparent, reproducible security
standards for open agent ecosystems. SARS captures instruction-layer and
multi-agent risk dimensions that CVSS does not; CVSS provides an internationally
recognized standard; and the ClawHub dual-view exposes gaps in official
marketplace safety reviews. With 1{,}299 evaluations and a growing corpus, \svb{}
is at once a benchmark, a dataset, and an actionable evaluation service. The
platform is freely accessible at \url{\sburl}.

\appendix
\section{Reproducibility and Data Statement}
\label{app:repro}

\textbf{Models and versions.}
Default leaderboard judge: \textsc{Qwen2.5-14B-Instruct}. Benchmark judges:
\textsc{Qwen2.5-32B}, \textsc{Llama-3.1-7B}, \textsc{Llama-3.1-70B},
\textsc{Llama-3.2-3B-Instruct}, \textsc{Mixtral-8x7B}.

\textbf{Decoding and limits.}
Temperature $0.2$, \texttt{top\_p} $0.9$, single judge and single prompt template
per evaluation across all sub-tasks.
\textbf{Artifact formats and preprocessing.}
Input is the skill artifact (\texttt{SKILL.md}), parsed into type-tagged segments
(\texttt{<instructions>}),
normalized, concatenated in fixed order, and truncated code/config-last when over
budget (logged).

\textbf{Condensed judge prompt template.}
The production judge uses the fixed template below (abridged: the 15 category
descriptions and per-metric scoring tables are summarized rather than reproduced
in full). Locate the prompt within the full codebase available at \url{https://github.com/supreme-lab/SkillVetBench/tree/master}.

\begin{lstlisting}[style=prompt]
[ROLE] You are a senior cybersecurity researcher specializing in AI agent
systems. Evaluate one agent-skill artifact (SKILL.md) for security risk.

[TASK] Check the skill against all 15 vulnerability categories:
  Code-level (1-12): Command/Shell Injection; Unsafe File Operations; Remote Code Execution;
  Data Exfiltration; Dependency/Supply Chain; Prompt Injection; Privileg Escalation;
  Credential/Secret Exposure; Indirect/Embedded Injection; Scope Creep / Over-privileged Tool Use; Insecure Deserialization; Log/Output Injection.
Agentic (13-15): Memory Poisoning & Persistence; Agentic State Manipulation; Multi-Agent / Subagent Attacks. Flag dangerous code blocks, NL instructions that run commands or treat external/retrieved content as instructions, install/download steps, over-broad access, and unvalidated memory or subagent writes.

[RUBRIC: SARS] Score five dimensions as integers 0-3 (see Table 1):
  IFR Instruction Fidelity Risk | DG Data Gravity | AI Action Irreversibility | BR Blast Radius | CA Chain Amplification.

[RUBRIC: CVSS v4.0] Assign the base-metric VECTOR for the worst/aggregate finding using v4.0 keys only (AV:N, AC:L fixed for agentic skills):
  AT in {N,P}; PR in {N,L,H}; UI in {N,P,A}; VC,VI,VA,SC,SI,SA in {H,L,N}; E in {A,P,U,X}.
(The numeric score is computed from this vector by the standard FIRST
function, not by the model.)

[OUTPUT] Return ONLY valid JSON (no fences, no preamble): {
    "skill_name", "overall_risk" in {CRITICAL,HIGH,MEDIUM,LOW,NONE},
    "is_vulnerable", "vulnerability_count",
    "cvss_metrics": {AT,PR,UI,VC,VI,VA,SC,SI,SA,E}, "sars_metrics": {IFR,DG,AI,BR,CA},
    "vulnerabilities": [ {id:"SKV-NNN", category, title, severity, affected_content, explanation, attack_scenario, remediation} ], "executive_summary", "skill_purpose_analysis", "dangerous_patterns_found": [...], "safe_patterns_noted": [...], "remediation_priority"
  }
\end{lstlisting}


\textbf{Scoring rubric.}
The SARS ordinal rubric used by the judge is reproduced verbatim in
Table~\ref{tab:sars_rubric}; the taxonomy definitions are in
Table~\ref{tab:taxonomy}. The CVSS numeric score is computed from the
judge-assigned vector by the standard FIRST scoring function~\cite{cvssv4}.

\textbf{Dataset construction and labeling.}
Detection results (Section~\ref{sec:llmjudge}) use the labeled 100-skill set
(78 malicious + 22 benign) from the companion benchmark~\cite{hossain2026svb};
the live leaderboard corpus comprises 1{,}299 skills drawn from the
ClawHub/OpenClaw marketplace.

\textbf{Data statement and snapshot.}
Per-skill reports, SKV findings, and CSV export are publicly available via the
Hugging Face Space at \url{\sburl}. Because the leaderboard is live, all
statistics in Section~\ref{sec:stats} are anchored to the \textbf{24 May 2026}
snapshot; current numbers on the live Space may differ.

\end{document}